\documentclass[aps,twocolumn]{revtex4}

\usepackage{epsfig}

\begin{document}

\bibliographystyle{apsrev}

\title{Solid immersion lens--enhanced micro--photoluminescence: principle and applications}
\author{S.~Moehl}
\author{H.~Zhao}
\author{B.~Dal~Don}
\author{S.~Wachter}
\author{H.~Kalt}
\affiliation{Institut f\"ur Angewandte Physik, Universit\"at
Karlsruhe, D-76131 Karlsruhe, Germany}

\begin{abstract}
We demonstrate the combination of a hemispherical solid immersion
lens with a micro--photoluminescence set--up. Two advantages
introduced by the SIL, an improved resolution of 0.4 times the
wavelength in vacuum and a 5~times enhancement of the collection
efficiency, make it an ideal system for spatially resolved
spectroscopy applications. The influence of the air gap between
the SIL and the sample surface is investigated in detail. We
confirm the tolerance of the set--up to an air gap of several
micrometers. Such a system is proven to be ideal system in the
studies of exciton transport and polarization dependent single
quantum dot spectroscopy.
\end{abstract}

\maketitle

\section{Introduction}
Photoluminescence (PL) is one of the most important methods to
investigate electronic states in semiconductors. Since
semiconductor nanostructure becomes increasingly important in
optoelectronic applications, the PL has been developed to local
spectroscopy. To realize spatial resolution, one needs local
excitation or local detection or both. In far--field optics, the
resolution is limited by diffraction. In a common
micro-photoluminescence ($\mu$-PL) system, the achieved resolution
is about 1~$\mu$m. This limitation of the resolution can be
overcome by working in the near--field regime, where the
diffraction limit is not yet established. Scanning near-field
optical microscopy (SNOM), designed according to this idea,
realized a resolution of $\sim$ 100~nm. Despite of this high
resolution, several inherent as well as technical
problems\cite{apl1791} limit the performance of SNOM.

Another way to deal with the diffraction limitation is to increase
the refractive index of the media around the sample, i.e., to
increase the effective numerical aperture ($\mathrm{NA}_{eff}$) of
the optical system. This can be realized by immersing the sample
into oil or put a tiny lens, named solid immersion lens (SIL), on
the surface of the sample. In the case of semiconductor
spectroscopy, the latter is more suitable since SIL is easily to
be dealt with and there is no risk to contaminate the sample.

During the last decade, SIL has been used in solid immersion
microscope\cite{apl2615} and (magneto--)optical data
storage\cite{apl388} for high spatial resolution or high storage
density, respectively. Recently, SIL has also been introduced in
spatially resolved pump--probe
experiments.\cite{apl1791,jm194523}. By including a superspherical
SIL\cite{bokborn}(s--SIL) in a $\mu$--PL system, an improved
spatial resolution at room temperature\cite{jjapl962} as well as
low temperatures\cite{apl635} has been demonstrated by PL imaging
measurements of GaAs quantum well (QW). The high spatial
resolution has allowed to study carrier migration under
global\cite{apl635} or local\cite{apl2965} excitation conditions.
Besides, s--SIL has also been used in a scanning $\mu$--PL setup
to investigate exciton localization in GaAs QW
\cite{prl2652,pssb505}.

Up to now, only s--SIL has been applied in $\mu$--PL system. But,
the thickness of an s--SIL is designed for one particular
wavelength since the incident parallel beam is focused at the
distance $r(1+1/n_{\mathrm{SIL}})$ away from the top of the
s--SIL, where $r$ is the radius of the SIL and $n_{\mathrm{SIL}}$
is the refractive index of the SIL material. Consequently, the
focus of an s--SIL is wavelength-dependent since
$n_{\mathrm{SIL}}$ depends on the wavelength of light, $\lambda$.
On the contrast, a hemispherical SIL (h--SIL) is universal for any
wavelength. In a PL experiment, one typically deals with different
wavelengths for excitation and detection. Thus, although an s-SIL
can improve the resolution $n_{\mathrm{SIL}}^{2}$ times while an
h-SIL can only improve it $n_{\mathrm{SIL}}$ times, the latter is
more appropriate for PL studies.

In this paper, we demonstrate for the first time the combination
of an h-SIL with a confocal $\mu$-PL system. We confirm the
realized spatial resolution of $0.4\lambda$ by introducing the
h-SIL. We also find an enhancement of collection efficiency of
about 5 times, which is consistence with a theoretical estimation.
We discuss in detail the influences of an air gap between the SIL
and the sample surface on the resolution and collection
efficiency. We demonstrate that an air gap of several micrometers
can be tolerated in a system with $\mathrm{NA}_{eff}<1$. Finally,
some applications of this system to semiconductor spectroscopy are
discussed, with an emphasis on its advantages over a SNOM system.

\section{Solid immersion lens--enhanced micro--photolumine-scence}
\subsection{Experimental Setup}
Figure~1 shows schematically the SIL--enhanced $\mu$--PL system.
The excitation source can be a continuous wave (cw) or pulse
laser. The laser beam is expanded to fit the diameter of
objective, then reflected by a beam--splitter and focused on the
sample surface through a microscope objective (20~$\times$,
numerical aperture $\mathrm{NA}_{Obj}=0.4$). The same objective is
used for collecting the PL from the sample. The signal passes the
beam-splitter and a tube lens and is then focused to the image
plane of the microscope. A set of pinholes, with different sizes,
is installed in the image plane to select the detection area. By
scanning the pinhole in the image plane, one can detect
luminescence from different positions on the sample surface. For
cw measurements, the PL is recorded by a double-grating
monochromator and a cooled CCD camera, with a spectral resolution
of 30~$\mu$eV. For time--resolved measurements, a streak camera
with temporal resolution of 2~ps is used in combination with a CCD
camera with photon counting.
\begin{figure}[t]
    \psfig{figure=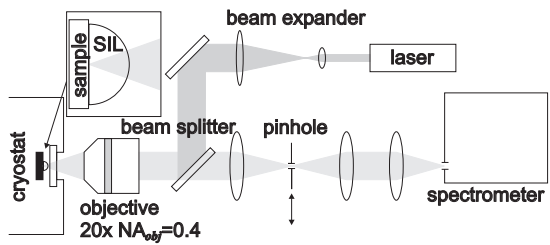,width=.4\textwidth}
    \caption{
    Experimental setup of the SIL-enhanced $\mu$-PL. The
    inset shows the SIL--sample configuration.
    }
\end{figure}

The h--SIL, made of Zr$\mathrm{O}_{2}$ with
$n_{\mathrm{SIL}}=2.16$ at $\lambda=600$~nm, is adhesively fixed
to the sample surface. The sample with the SIL is vertically
mounted in a helium--flow cryostat. The SIL can be used in the
temperature range of 6$\sim$300~K for an unlimited number of
cooling cycles. The diameter of the SIL is chosen to be 1~mm,
which is large enough for giving an enough working area for
spectroscopy and being handled without any special equipment, and
still small enough to be sticked on the sample adhesively even in
vertical configuration.

\subsection{Spatial Resolution}
In a far--field optical system, the spatial resolution is limited
by diffraction of the light. For a plane wave incident light, the
half width at half maximum (HWHM) of the Airy pattern is given by
\begin{equation}
  \mathrm{HWHM}=\frac{0.26\lambda}{n\mathrm{NA}_{obj}}.
\end{equation}
Here, $n$ represents the refraction index of the media around the
sample. Without SIL, $n\approx1$ (air) and with the h--SIL we have
$n=n_{\mathrm{SIL}}=2.16$. Thus, by introducing the h--SIL, we can
improve the resolution by more than 2~times. In order to confirm
the achieved resolution, we install the SIL onto an arbitrary flat
sample and focus the incident laser beam of a He--Ne laser on the
sample surface (i)underneath and (ii)outside the SIL,
respectively. We measure two--dimensional intensity maps of the
laser spots in both case, as shown in Fig.~2(a) with the same
color encoding. The length--scale collaboration in these maps is
obtained by imaging an optical grating with known parameters. The
spatial intensity profiles in Fig.~2(b) are obtained by taking a
line--scan across the laser spots. As expected, the profile with
SIL (i) is about 2~times narrower than that obtained without SIL
(ii). In (i) the realized spatial resolution (HWHM of the laser
spot) is $0.4\lambda$ (corresponding to 260~nm for the He--Ne
laser (633~nm)) in contrast to $0.8\lambda$ for (ii). We note that
the achieved values of HWHM in both cases are larger than that
calculated from Eq.~1. This is consistent with theoretical
calculation\cite{prsl358}, and can be attributed to the Guassian
profile, rather than plane wave, of the laser beam\cite{rpp427}
used in the experiment and the high NA\cite{prsl358} of the
system.
\begin{figure}[t]
    \psfig{figure=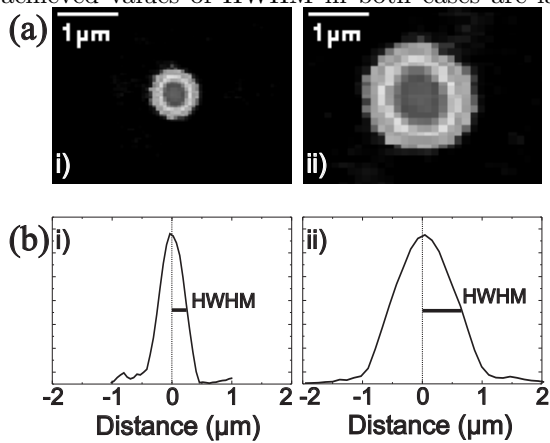,width=.4\textwidth}
    \caption{Intensity maps (a) and cross-sections (b) of the focused
    laser on an arbitrary sample. The width of the spot obtained with
    SIL (i) is $n_{\mathrm{SIL}}$ times narrower than that without SIL (ii).
    \label{fig:henespot}}
\end{figure}

In a confocal $\mu$-PL system, the resolution can be further
improved by introducing a pinhole with a suitable size to the
image plane of the microscope.\cite{rpp427} In the followings, we
present a quantitative analysis of this further improvement. The
illumination function of the laser excitation can be described by
a Guassian function,
\begin{equation}
    i_{\mathrm{ill}}(q)=\mathrm{exp}(-2\frac{q^{2}}{w_{\mathrm{laser}}^{2}}).
\end{equation}
Here, $q$ is the coordinate in the focal plane, i.e., sample
surface, and $w_{\mathrm{laser}}$ is the spot radius at $1/e^{2}$.
The detection function $i_{\mathrm{det}}$ can also be described by
Guassian function, but with a different radius $w_{\mathrm{lumi}}$
since generally the wavelength of the luminescence is different
from that of the excitation laser in a PL experiment. The
transmission function of the pinhole is
\begin{equation}
 t_{\mathrm{p}}(q)=\mathrm{rect}(\frac{q}{q_{0}})=\left\lbrace {1\quad | q |<q_{0}\atop 0\quad | q
 |>q_{0}}
 \right.
\end{equation}
with $q_{0}$ the radius of the pinhole image. Thus, the detection
probability, i.e., the probability of a photon emitted at point
$p$ transmit the pinhole thus be detected, is given by
\begin{widetext}
\begin{equation}
    c(q)=t_{\mathrm{p}}(q)\ast i_{\mathrm{det}}(q)
        =\int_{0}^{q_{0}}\! \! q'dq'\int_{0}^{2\pi}\! \!
     d\Phi'\mathrm{exp}(-2\frac{q^{2}-2qq'cos\Phi'+q'^{2}}{w_{\mathrm{det}}^{2}})\\.
\end{equation}
\end{widetext}
The confocal acceptance function (CAF) is then given by
\begin{equation}
   p_{\mathrm{conf}}(q)=i_{\mathrm{ill}}(q)\cdot c(q).
\end{equation}

Based on the above analysis, we calculate the $p_{conf}$ of our
SIL--enhanced $\mu$--PL system. Figure 3 shows the calculated HWHM
of the CAF, which defines the confocal resolution, as function of
the pinhole size. The horizontal line represents the resolution
obtained without pinhole. We find that a pinhole of 60~$\mu$m has
no effect on the resolution, but decreasing the pinhole size from
that value the resolution is enhanced. Below 10~$\mu$m, the
enhancement is saturated while further decrease the pinhole size.
\begin{figure}[t]
    \psfig{figure=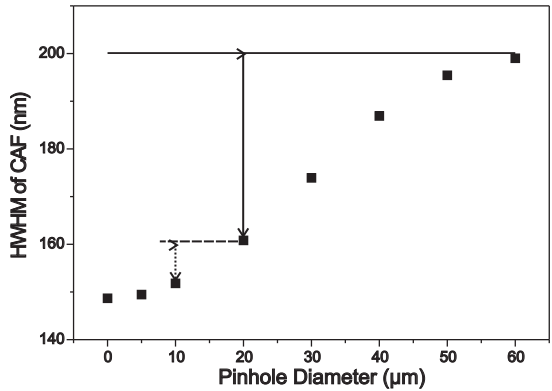,width=.4\textwidth}
    \caption{Calculated HWHM of the confocal acceptance function (CAF)
     as function of the pinhole diameter. The horizontal full line represents
     the HWHM obtained with an infinite large pinhole. In the calculation, the
     excitation and detection wavelengths are 476.5~nm and 529~nm, respectively, which are
     consistent with the experimental conditions.}
\end{figure}

In order to confirm that the enhancement of resolution by the SIL
and pinhole can be achieved in a realistic PL measurement, we
measure the spectra from a ZnCdSe/ZnSe quantum dot sample with
different SIL--pinhole configurations. In this sample, a ZnCdSe
layer with a thickness of 2.9 monolayers is embedded between two
ZnSe barriers, including Cd--rich quantum dots with an average
size of about 10~nm. The spectrally sharp lines in the spectrum
correspond to excitonic transitions in individual dots. The
variations in size, shape and composition of these dots lead to a
wide spectral distribution of the lines. Hence, in a macroscopic
PL spectrum (not shown here), one observes a broad smooth emission
band due to the large number of contributing dots. When decreasing
the detection area, hence the number of the dots, individual sharp
lines can be resolved on top of the unresolved smooth background.
The resolved part becomes more and more pronounced with decreasing
detection area. Eventually, once the detection area is small
enough, the unresolved background will disappear. Such a kind of
sample, with suitable dot density, can be used to prove
qualitatively the enhancement of spatial resolution by introducing
the SIL.

Figure~4 shows four spectra detected at a sample temperature of
6~K with different SIL--pinhole configurations, i.e., without SIL
and pinhole (A), with SIL but without pinhole (B), with SIL and a
20~$\mu$m diameter pinhole (C), with SIL and a 10~$\mu$m diameter
pinhole, respectively. The sample is excited by the 476.5~nm line
of an Ar--ion laser. All spectra are composed of a resolved and an
unresolved part, but the resolved sharp lines in the spectrum is
more pronounced as we go from (A) to (D). We fit the background by
a Gaussian function in order to separate the resolved and the
unresolved part. The choice of a Gaussian is legitimate because of
the inhomogeneous distribution of a large number of quantum dots
contributing to the spectra. For each spectrum, we calculate the
ratio, $R$, of the spectrally integrated intensities of the
resolved part to the unresolved smooth background. Such a ratio
increases with enhancing the resolution of the system, as we
discussed above. From Fig.~4 we obtain an increase of $R$ by
30~$\%$ by introducing the SIL (compare 0.109 of A to 0.143 of B).
By introducing a 20~$\mu$m pinhole, $R$ is further increased by
20~$\%$ (0.171 of C). In case D, a 10~$\mu$m pinhole is used
instead of the 20~$\mu$m one. But we don't find a further increase
of $R$ (0.170 of D). This is consistent with our analysis
discussed above. As shown in Fig.~3, the enhancement of the
resolution introduced by changing a 20~$\mu$m pinhole to a
10~$\mu$m one (vertical dots) is much smaller than that from
non--pinhole to 20~$\mu$m pinhole (vertical full line). In
practice, the signal level drops seriously as decreasing the
pinhole size from 20~$\mu$m, and the alignment becomes more
difficult. Thus, a pinhole size of 20~$\mu$m is the optimal choice
in our system.
\begin{figure}[t]
    \psfig{figure=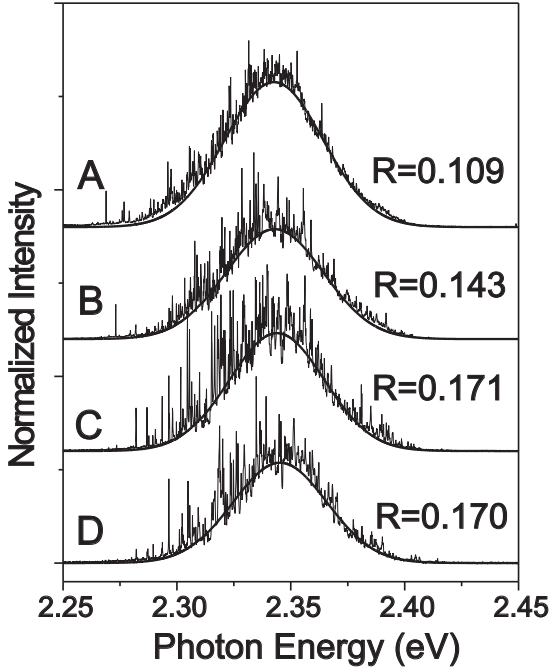,width=.4\textwidth}
    \caption{$\mu$-PL spectra of a ZnCdSe/ZnSe quantum island sample
    measured with different configurations. A: without SIL and pinhole;
    B: with SIL but without pinhole; C: with SIL and 20~$\mu$m pinhole;
    D: with SIL and 10~$\mu$m pinhole}
\end{figure}

\subsection{Collection Efficiency}
In a PL experiment, only a part of luminescence from the sample
can be collected due to the reflection losses and the finite size
of the optics. The collection efficiency of a spectroscopy system
is of crucial importance, especially in the cases of low
excitation conditions or low signal level. Since it is operated in
the far field, the collection efficiency of a $\mu$--PL system is
much higher than that of typical near--field systems. By
introducing a SIL into a $\mu$--PL system, the collection
efficiency can be further improved. By comparing the luminescence
intensities measured with SIL and without SIL, we find the
enhancement of collection efficiency of our system by the SIL is
about 5 times, depending mainly on the cleaning process of both
the sample and the SIL.

Here, we present a quantitative analysis on the enhancement of the
collection efficiency introduced by using the SIL. Since the
$n_{\mathrm{SIL}}$ is smaller than the refraction index of the
sample, $n_{\mathrm{samp}}$, the SIL has the property to reduce
the reflection losses, i.e., enhance the transmission of both the
luminescence and the laser. The enhancement of the collection
efficiency by this factor, $k_\mathrm{T}$, can be calculated by
using Fresnel formula. Figure~5(a) shows the configurations for
our calculation of the $k_\mathrm{T}$ by comparing the
transmissions when the SIL is used (i) or not (ii). In case (i),
since the light enters perpendicularly on the top of the SIL, the
transmission coefficient of the intensity is given by $
4n_{\mathrm{SIL}}/(1+n_{\mathrm{SIL}})^{2}$ for all rays.
\begin{figure}[t]
    \psfig{figure=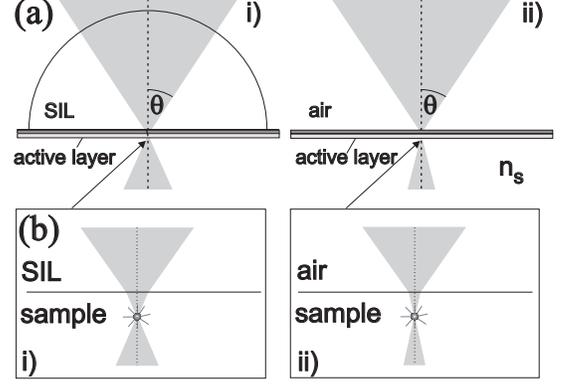,width=.4\textwidth}
    \caption{Schematic drawing of the SIL-sample configuration.
    The enhancement of collection efficiency is explained by higher
    transmission (a) and larger collection angle in the sample (b).\label{fig:colleff}}
\end{figure}
However, when entering the sample, the transmission coefficient
depends of the angle of incidence and the polarization of the ray.
This angle dependency is weak in the range of angles given by the
microscope objective. Thus, for average, we calculate for each
polarization the transmission of a ray with an angle to the
optical axes of $\theta/2$. Furthermore, the transmissions of s
and p polarizations are averaged to get the total transmission.
Considering the reflection losses of both the laser and the
luminescence, we get an enhancement factor $k_\mathrm{T}=1.2$.

Beside this transmission enhancement, the SIL can enlarge the
collection angle of the $\mu$--PL system, as shown in Fig.~5(b).
The solid angle outside of the sample is independent of whether
the SIL is used (i) or not (ii) and is directly given by
$\mathrm{NA}_{obj}$. However, the solid angle inside the sample
increases when the SIL is introduced. This is due to the smaller
refraction of the light at the sample surface since the material
on top of the sample having now a refractive index higher than
that of air. As a result, a point source emitting light in all
directions as shown in Fig.~5(b) experiences a larger solid angle
in which the emitted photons can be collected by the objective. A
ray emitted outside of this angle will miss the objective and not
contribute to the signal, even if its angle to the optical axis is
smaller than the critical angle of total internal reflection. In
the approximation that the photons are emitted homogeneously in
all directions, the enhancement of collection efficiency due to
the larger PL collection angle, $k_\mathrm{\Omega}$, is given by
the ratio of the solid angles $\Omega_{i}$ with SIL to
$\Omega_{ii}$ without SIL:
\begin{widetext}
\begin{equation}
 k_\mathrm{\Omega}=\frac{\Omega_{\mathrm{SIL}}}{\Omega_{\mathrm{air}}} \approx
 \frac{1-\cos \left( \frac{n_{\mathrm{SIL}}}{n_{\mathrm{samp}}}
 \sin\theta \right)}{1-\cos\left( \frac{1}{n_{\mathrm{samp}}}
 \sin\theta \right)} \approx \frac{1-\left( 1-\frac{1}{2}
 \left( \frac{n_{\mathrm{SIL}}}{n_{\mathrm{samp}}} \right)^{2}
 \sin^{2}\theta \right)}{1-\left( 1-\frac{1}{2}
 \left( \frac{1}{n_{\mathrm{samp}}} \right)^{2}\sin^{2}\theta
 \right)}=n_{\mathrm{SIL}}^{2}.
\end{equation}
\end{widetext}
Thus, the total enhancement of the collection efficiency by SIL is
\begin{equation}
 k_\mathrm{total}=k_\mathrm{T} \cdot k_\mathrm{\Omega} \approx
 k_\mathrm{T} \cdot n_{\mathrm{SIL}}^{2}.
\end{equation}
In our set--up, we have $k_\mathrm{T}=1.2$,
$k_\mathrm{\Omega}=4.8$ so $k_\mathrm{total}=5.76$. This
calculated value is consistent with out experimental results.

\section{Influence of air gap}
In the analyses of the previous section, we assume that the SIL is
ideally attached on the sample surface. In a realistic experiment,
there exists an air gap between the flat surface of the SIL and
the sample surface due to the fluctuations of both surfaces as
well as due to particles between them. In this section, we discuss
the influence of such an air gap on the resolution and collection
efficiency of the SIL--enhanced $\mu$--PL system.

As discussed above, the $\mathrm{NA}_{eff}$, of a SIL--enhanced
$\mu$--PL system is determined by the $\mathrm{NA}_{obj}$ and
$n_{\mathrm{SIL}}$, i.e., $\mathrm{NA}_{obj} \cdot
n_{\mathrm{SIL}}$ for h--SIL and $\mathrm{NA}_{obj} \cdot
n_{\mathrm{SIL}}^{2}$ for s--SIL. The influence of the air gap on
the resolution depends strongly on whether $\mathrm{NA}_{eff}>1$
or not. In a system with $\mathrm{NA}_{eff}>1$, near--field
coupling is required. Theoretical analysis shew that even an air
gap with a thickness of one fifth of the wavelength can
deteriorate the resolution seriously.\cite{jap6923} In contrast, a
system with $\mathrm{NA}_{eff}<1$ is still in far--field regime,
and it has been shown theoretically that an air gap of several
micrometers doesn't influence the resolution.\cite{inp56}. In our
set--up, we have $\mathrm{NA}_{eff}=0.87<1$ thus far--field
coupling. To check the influence of air gap on the resolution of
our system, we load the SIL on the sample with and without
cleaning procedure, respectively. In the latter case, an air gap
of several micrometers is anticipated (we will prove this fact
later). We focus the laser beam on the sample surface, and in both
cases we get the same size of the laser spots. We even load the
SIL on the surface of an optical grating without any special
treatment and obtain the same resolution. Thus we confirm that in
a system with $\mathrm{NA}_{eff}<1$, an air gap of several
micrometers has no influence on the resolution.

Generally, an air gap introduces additional reflection losses
between the sample and the SIL, thus reduces the collection
efficiency. In a near--field system with $\mathrm{NA}_{eff}>1$,
the collection efficiency can be deteriorated seriously by an air
gap of several hundreds nanometers, i.e., comparable to the light
wavelength.\cite{inp56} In contrast, a system with
$\mathrm{NA}_{eff}<1$ is anticipated to be more robust due to the
far--field feature. In order to investigate the tolerance of our
system to the air gap, we load the SIL on a ZnCdSe/ZnSe quantum
dot sample without any cleaning procedure. By comparing the
spectra measured beneath or outside the SIL at a sample
temperature of 6~K, we find an \emph{enhancement} of collection
efficiency by a factor of 2.

The explain the observed enhancement, we calculate the collection
efficiency of the system with an air gap. Figure~6 shows the
configuration of the SIL, air gap and the sample. As measuring the
spectrum, we focus on the sample surface. The path of the ray is
shown as the solid line in Fig.~6. We can also find the focus of
the flat surface of the SIL, as shown as the dots. Since the
images of both surfaces are clearly observable, the distance
between these two foci, $d$, can be measured accurately. In this
experiment, we have $d$=40~$\mu$m. By some simple geometrical
considerations, we deduce the thickness of the air gap to be
5~$\mu$m from the measured $d$. Based on Fig.~6, we calculate the
collection efficiency of this configuration by the method
discussed in the previous section. We obtain $k_\mathrm{T}=0.55$,
$k_\mathrm{\Omega}=4.27$ so $k_\mathrm{total}=2.36$. The
calculation is well consistent with the experiment. We note that
the deterioration of enhancement from 5.76 to 2.36 by the 5~$\mu$m
air gap originates mainly from the increasing of the reflection
losses ($k_\mathrm{T}$ drops from 1.2 to 0.55). The enhancement
due to the enlarged collection angle, $k_\mathrm{\Omega}$, is not
sensitive to the presence of the air gap.
\begin{figure}[t]
    \psfig{figure=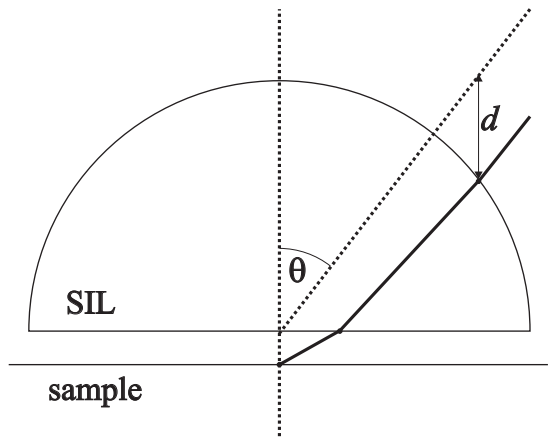,width=.4\textwidth}
    \caption{Schematic drawing of the SIL-sample configuration with an air gap.
    }
\end{figure}

Thus, we prove the tolerance of the SIL--enhanced $\mu$--PL system
to an air gap of several micrometers. This feature is quite
difference from a near--field system with $\mathrm{NA}_{eff}>1$.
In a typical measurement, there exists an air gap of about 1~$\mu$
thick between the SIL and the sample surface after a regular
cleaning procedure. The enhancement factor of collection
efficiency is about 5 in our experiments, as mentioned above. In
principle, by increasing the $\mathrm{NA}_{obj}$ or
$n_{\mathrm{SIL}}$, or using a s--SIL, one can further improve the
resolution of a SIL--enhanced $\mu$--PL system. But, if the
$\mathrm{NA}_{eff}$ is increased beyond 1, the near--field regime
is reached, and the tolerance to the air gap drops seriously. In
this sense, our choice of $\mathrm{NA}=0.864$ is a good compromise
between the enhancement of spatial resolution and collection
efficiency as well as the feasibility in practical operations.
Further, it is worth noting that such a system can be used even
for samples with rough surfaces with fluctuations of several
micrometers.

\section{Applications}
Generally, the SIL--enhance $\mu$--PL system described above can
be applied to any semiconductor spectroscopy experiments whenever
the spatial resolution is needed and the sample has a relative
flat surface. In some cases, this system is even more suitable
than any other spatially resolved methods. We present in the
following two kinds of applications where the SIL--enhanced
$\mu$--PL system shows its unique advantages.

\subsection{Exciton Transport in Quantum Wells}
The exciton in-plane transport process is an important part of
exciton dynamics in QWs. Due to the continues miniaturization of
electronic and optical devices thus the increasing importance of
nanostructures, the transport has to be understood on a length
scale comparable to the light wavelength. The resolution of a
conventional $\mu$--PL system, about 1~$\mu$m, is not enough for
this kind of studies. In SNOM, one can achieve a resolution of
$\sim$ 100~nm by using the coated tip. But the collection
efficiency is very poor. By using a uncoated tip, the collection
efficiency can be improved, but simultaneously the resolution
drops.\cite{apl76203} Furthermore, a SNOM experiment has another
disadvantage for transport measurement: one cannot detect
spatially resolved spectra from positions outside the excitation
spot. Thus, one can only study the transport process indirectly.

In contrast, the SIL--enhanced $\mu$--PL can be used to
investigate the transport behavior in a rather direct way with
sub-$\mu$m resolution. By scanning the pinhole in the image plane
of the objective, one can detect luminescence from positions which
are different from the position where the sample is locally
excited. This enables one to get the spatial profile of the
luminescence intensity, thus the spatial distributions of the
carrier density. The field of view of the SIL\cite{inp56}
(35~$\mu$m in our setup) is far beyond the typical transport
length of carriers. In Fig.~7, we show an example of the
\begin{figure}[t]
    \psfig{figure=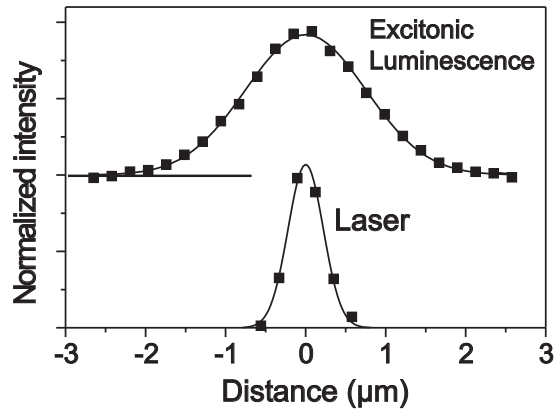,width=.4\textwidth}
    \caption{Spatial profile of the excitonic
           luminescence of a ZnSe/ZnSSe MQW (upper) and the
           corresponding excitation laser spot (lower).
    }
\end{figure}
spatial profiles of the luminescence intensity measured for a
ZnSe/ZnSSe multiple quantum well at 6~K. The excitation laser used
for this measurement is a tunable cw Ti:Sapphire laser pumped by
an Ar--ion laser and frequency--doubled using a BBO crystal.
Comparing with the profile of the laser excitation spot, measured
in the same pinhole--scanning, the spatial distribution of the
luminescence is significantly wider due to exciton in-plane
transport. The improved resolution enables us to access the hot
exciton transport regime.\cite{apl801391}

This kind of direct transport measurement can also be performed
with time resolution by using a short pulse Ti:Sapphire laser and
a synchroscan streak camera and CCD with photon counting. In this
configuration, one can detect the time evolution of the spatial
profile of luminescence, as shown in Fig.~8(b) for a ZnSe/ZnMgSSe
multiple QW. The combination of the 200~nm spatial resolution with
the 2~ps temporal resolution enables one to investigate more
details of the exciton transport. We note that for the
time-resolved measurements, the high collection efficiency
introduced by the SIL is particularly crucial, not only because of
the signal losses at the pinhole but also because signal intensity
spread due to the time resolution.
\begin{figure}[t]
    \psfig{figure=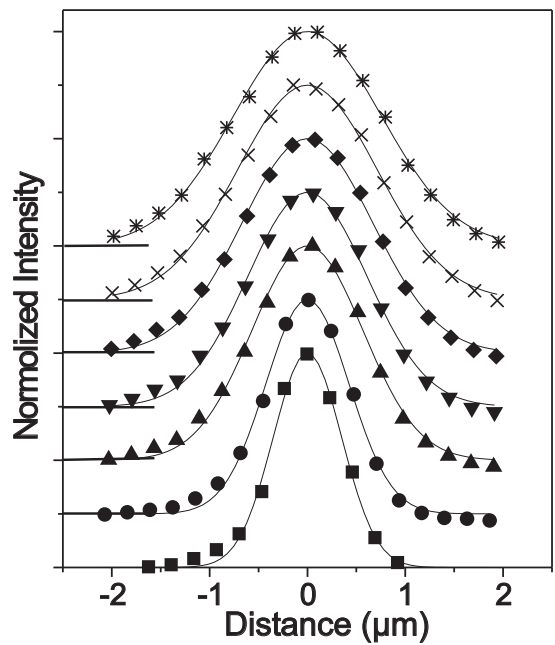,width=.4\textwidth}
    \caption{Time evolution of the spatial profile of the
           exciton luminescence. The ZnSe/ZnMgSSe MQW sample is
           excited by a 150~fs laser pulse. The times for the curves are (from bottom to top)
           43, 64, 100, 180, 260, 340 and 420~ps.\label{fig:znsespec}}
\end{figure}

\subsection{Single Quantum Dot Spectroscopy}
Investigation on properties of individual quantum dot requires
single dot spectroscopy. This can be achieved by small dot density
(e.g. by nano--aperture or mesa) or high spatial resolution (e.g.
SNOM). In SIL--enhanced $\mu$--PL, single dot spectroscopy can
also be achieved for samples with low quantum dot density. In such
a system, the choice of dots is more flexible. Since there is no
patterning required, it is a non--destructive method.

The high spatial and spectral resolutions of the SIL--enhance
$\mu$--PL system enable us to detect isolated narrow lines from
single quantum dot undisturbed by the luminescence from other
dots. Figure~9 shows a spectrum of a GaAs/AlAs superlattices. In
this kind of samples, quantum dots are formed due to the interface
fluctuations.\cite{s27387}. A He--Ne laser is used for excitation,
with an excitation density of 0.5 W/$\mathrm{cm^{2}}$. By using
the SIL and the 20~$\mu$m pinhole, isolated sharp lines from
individual quantum dots are well resolved.
\begin{figure}[t]
    \psfig{figure=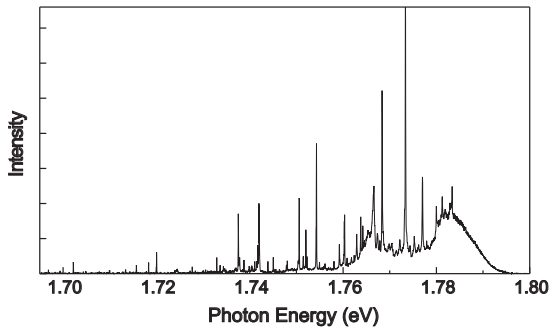,width=.4\textwidth}
    \caption{SIL--enhance $\mu$--PL spectrum of a GaAs/AlAs superlattices measured at
    a sample temperature of 6~K. The excitation source is a He--Ne laser, at an
    excitation intensity of 0.5 W/$\mathrm{cm^{2}}$. Isolated sharp lines from quantum dots formed by interface
    fluctuations are well resolved.
    }
\end{figure}

Furthermore, we confirm that the polarization information of the
luminescence, which is of crucial importance e.g. in the
investigations of spin dynamics, is preserved in this set--up. In
Fig.~10 we show the spectra of a single quantum dot in a quantum
well of 6.5 monolayers ZnCdTe embedded in ZnTe barriers. The
sample is excited with the 488~nm line from the cw Ar--ion laser.
To determine the polarization of luminescence, a linear polarizer
is placed in the detection path. The spectra show a doublet
structure composed of two lines which are linearly polarized along
two orthogonal directions. Such line doublets are ascribed to
fine--structure splitting of excitons in asymmetric quantum
dots.\cite{jcg742} Our measurement demonstrates that SIL can be
applied to $\mu$-PL when the polarization of the light is of
interest. It is typically very difficult to extract the
polarization from other spatially resolved techniques like
near-field spectroscopy.\cite{apl3929}
\begin{figure}[t]
    \psfig{figure=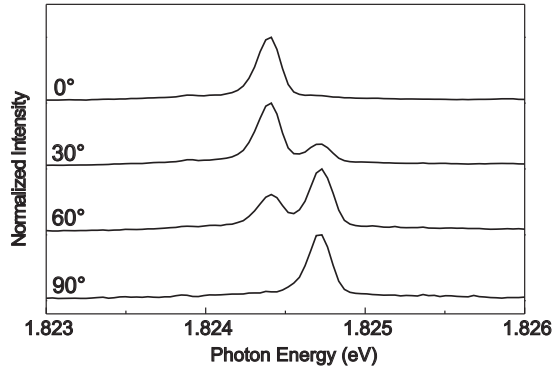,width=.4\textwidth}
    \caption{SIL--enhanced $\mu$--PL spectra of a single ZnCdTe quantum dot embedded in a
    ZnTe matrix as function of the angle of a linear polarizer in the detection path.
    }
\end{figure}

\section{Summary}
We demonstrate for the first time the combination of a h--SIL with
a $\mu$--PL set--up. Two advantages introduced by the SIL, an
improved resolution of 0.4$\lambda$ and a 5~times enhancement of
the collection efficiency, make it an ideal system for spatially
resolved spectroscopy applications. We analyze the improvement of
the resolution by using the pinhole, and find the optimal pinhole
size. The influence of the air gap between the SIL and the sample
surface is investigated in detail. We confirm the tolerance of the
set--up to an air gap of several micrometers. Such a system is
proven to be ideal system in the studies of exciton transport and
polarization dependent single quantum dot spectroscopy.

\section{Acknowledgments}
We gratefully acknowledge the growth of the samples by the group
of M. Heuken (Aachen) and the group of D. Hommel (Bremen). This
work was supported by the Deutsche Forschungsgemeinschaft.


\begin{thebibliography}{10}
\expandafter\ifx\csname bibnamefont\endcsname\relax
  \def\bibnamefont#1{#1}\fi
\expandafter\ifx\csname bibfnamefont\endcsname\relax
  \def\bibfnamefont#1{#1}\fi
\expandafter\ifx\csname url\endcsname\relax
  \def\url#1{\texttt{#1}}\fi
\expandafter\ifx\csname
urlprefix\endcsname\relax\def\urlprefix{URL }\fi
\providecommand{\bibinfo}[2]{#2}
\providecommand{\eprint}[2][]{\url{#2}}

\bibitem{apl1791}
\bibinfo{author}{\bibfnamefont{M.}~\bibnamefont{Vollmer}},
  \bibinfo{author}{\bibfnamefont{H.}~\bibnamefont{Giessen}},
  \bibinfo{author}{\bibfnamefont{W.}~\bibnamefont{Stolz}},
  \bibinfo{author}{\bibfnamefont{W.~W.} \bibnamefont{Rühle}},
  \bibinfo{author}{\bibfnamefont{L.}~\bibnamefont{Ghislain}}, \bibnamefont{and}
  \bibinfo{author}{\bibfnamefont{V.}~\bibnamefont{Elings}},
  \bibinfo{journal}{Appl. Phys. Lett.} \textbf{\bibinfo{volume}{74}},
  \bibinfo{pages}{1791} (\bibinfo{year}{1999}).

\bibitem{apl2615}
\bibinfo{author}{\bibfnamefont{S.~M.} \bibnamefont{Mansfield}}
  \bibnamefont{and} \bibinfo{author}{\bibfnamefont{G.~S.} \bibnamefont{Kino}},
  \bibinfo{journal}{Appl. Phys. Lett.} \textbf{\bibinfo{volume}{57}},
  \bibinfo{pages}{2615} (\bibinfo{year}{1990}).

\bibitem{apl388}
\bibinfo{author}{\bibfnamefont{B.~D.} \bibnamefont{Terris}},
  \bibinfo{author}{\bibfnamefont{H.~J.} \bibnamefont{Mamin}},
  \bibinfo{author}{\bibfnamefont{D.}~\bibnamefont{Rugar}},
  \bibinfo{author}{\bibfnamefont{W.~R.} \bibnamefont{Studenmund}},
  \bibnamefont{and} \bibinfo{author}{\bibfnamefont{G.~S.} \bibnamefont{Kino}},
  \bibinfo{journal}{Appl. Phys. Lett.} \textbf{\bibinfo{volume}{65}},
  \bibinfo{pages}{388} (\bibinfo{year}{1994}).

\bibitem{jm194523}
\bibinfo{author}{\bibfnamefont{M.}~\bibnamefont{Vollmer}},
  \bibinfo{author}{\bibfnamefont{H.}~\bibnamefont{Giessen}},
  \bibinfo{author}{\bibfnamefont{W.}~\bibnamefont{Stolz}},
  \bibinfo{author}{\bibfnamefont{W.~W.} \bibnamefont{R{\"u}hle}},
  \bibinfo{author}{\bibfnamefont{A.}~\bibnamefont{Knorr}},
  \bibinfo{author}{\bibfnamefont{S.~W.} \bibnamefont{Kock}},
  \bibinfo{author}{\bibfnamefont{L.}~\bibnamefont{Ghislain}}, \bibnamefont{and}
  \bibinfo{author}{\bibfnamefont{V.}~\bibnamefont{Elings}},
  \bibinfo{journal}{J.~Microscopy} \textbf{\bibinfo{volume}{194}},
  \bibinfo{pages}{523} (\bibinfo{year}{1999}).

\bibitem{bokborn}
\bibinfo{author}{\bibfnamefont{M.}~\bibnamefont{Born}} \bibnamefont{and}
  \bibinfo{author}{\bibfnamefont{E.}~\bibnamefont{Wolf}},
  \emph{\bibinfo{title}{{P}rinciples of {O}ptics}}
  (\bibinfo{publisher}{Pergamon}, \bibinfo{address}{Oxford},
  \bibinfo{year}{1970}).

\bibitem{jjapl962}
\bibinfo{author}{\bibfnamefont{T.}~\bibnamefont{Sasaki}},
  \bibinfo{author}{\bibfnamefont{M.}~\bibnamefont{Baba}},
  \bibinfo{author}{\bibfnamefont{M.}~\bibnamefont{Yoshita}}, \bibnamefont{and}
  \bibinfo{author}{\bibfnamefont{H.}~\bibnamefont{Akiyama}},
  \bibinfo{journal}{Jpn. J. Appl. Phys.}
  \textbf{\bibinfo{volume}{\textnormal{Part2} 36}}, \bibinfo{pages}{L962}
  (\bibinfo{year}{1997}).

\bibitem{apl635}
\bibinfo{author}{\bibfnamefont{M.}~\bibnamefont{Yoshita}},
  \bibinfo{author}{\bibfnamefont{T.}~\bibnamefont{Sasaki}},
  \bibinfo{author}{\bibfnamefont{M.}~\bibnamefont{Baba}}, \bibnamefont{and}
  \bibinfo{author}{\bibfnamefont{H.}~\bibnamefont{Akiyama}},
  \bibinfo{journal}{Appl. Phys. Lett.} \textbf{\bibinfo{volume}{73}},
  \bibinfo{pages}{635} (\bibinfo{year}{1998}).

\bibitem{apl2965}
\bibinfo{author}{\bibfnamefont{M.}~\bibnamefont{Yoshita}},
  \bibinfo{author}{\bibfnamefont{M.}~\bibnamefont{Baba}},
  \bibinfo{author}{\bibfnamefont{S.}~\bibnamefont{Koshiba}},
  \bibinfo{author}{\bibfnamefont{H.}~\bibnamefont{Sakaki}}, \bibnamefont{and}
  \bibinfo{author}{\bibfnamefont{H.}~\bibnamefont{Akiyama}},
  \bibinfo{journal}{Appl. Phys. Lett.} \textbf{\bibinfo{volume}{73}},
  \bibinfo{pages}{2965} (\bibinfo{year}{1998}).

\bibitem{prl2652}
\bibinfo{author}{\bibfnamefont{Q.}~\bibnamefont{Wu}},
  \bibinfo{author}{\bibfnamefont{R.~D.} \bibnamefont{Grober}},
  \bibinfo{author}{\bibfnamefont{D.}~\bibnamefont{Gammon}}, \bibnamefont{and}
  \bibinfo{author}{\bibfnamefont{D.~S.} \bibnamefont{Katzer}},
  \bibinfo{journal}{Phys. Rev. Lett.} \textbf{\bibinfo{volume}{83}},
  \bibinfo{pages}{2652} (\bibinfo{year}{1999}).

\bibitem{pssb505}
\bibinfo{author}{\bibfnamefont{Q.}~\bibnamefont{Wu}},
  \bibinfo{author}{\bibfnamefont{R.~D.} \bibnamefont{Grober}},
  \bibinfo{author}{\bibfnamefont{D.}~\bibnamefont{Gammon}}, \bibnamefont{and}
  \bibinfo{author}{\bibfnamefont{D.~S.} \bibnamefont{Katzer}},
  \bibinfo{journal}{Phys. Stat. Sol. B} \textbf{\bibinfo{volume}{221}},
  \bibinfo{pages}{505} (\bibinfo{year}{2000}).

\bibitem{prsl358}
\bibinfo{author}{\bibfnamefont{B.}~\bibnamefont{Richards}} \bibnamefont{and}
  \bibinfo{author}{\bibfnamefont{E.}~\bibnamefont{Wolf}},
  \bibinfo{journal}{Proc. R. Soc. London} \textbf{\bibinfo{volume}{A253}},
  \bibinfo{pages}{358} (\bibinfo{year}{1959}).

\bibitem{rpp427}
\bibinfo{author}{\bibfnamefont{R.~H.} \bibnamefont{Webb}},
  \bibinfo{journal}{Rep. Prog. Phys.} \textbf{\bibinfo{volume}{59}},
  \bibinfo{pages}{427} (\bibinfo{year}{1996}).

\bibitem{jap6923}
\bibinfo{author}{\bibfnamefont{M.}~\bibnamefont{Baba}},
  \bibinfo{author}{\bibfnamefont{T.}~\bibnamefont{Sasaki}},
  \bibinfo{author}{\bibfnamefont{M.}~\bibnamefont{Yoshita}}, \bibnamefont{and}
  \bibinfo{author}{\bibfnamefont{H.}~\bibnamefont{Akiyama}},
  \bibinfo{journal}{J. Appl. Phys.} \textbf{\bibinfo{volume}{85}},
  \bibinfo{pages}{6923} (\bibinfo{year}{1999}).

\bibitem{inp56}
\bibinfo{author}{\bibfnamefont{G.~S.} \bibnamefont{Kino}}, in
  \emph{\bibinfo{booktitle}{Optical pulse and beam propagation}}
  (\bibinfo{publisher}{SPIE}, \bibinfo{address}{Washington},
  \bibinfo{year}{1999}), vol. \bibinfo{volume}{3609} of
  \emph{\bibinfo{series}{Proceedings of the SPIE}}, p.~\bibinfo{pages}{56}.

\bibitem{apl76203}
\bibinfo{author}{\bibfnamefont{G.}~\bibnamefont{von Freymann}},
  \bibinfo{author}{\bibfnamefont{D.}~\bibnamefont{L\"uer$\beta$en}},
  \bibinfo{author}{\bibfnamefont{C.}~\bibnamefont{Rabenstein}},
  \bibinfo{author}{\bibfnamefont{M.}~\bibnamefont{Mikolaiczyk}},
  \bibinfo{author}{\bibfnamefont{H.}~\bibnamefont{Richter}},
  \bibinfo{author}{\bibfnamefont{H.}~\bibnamefont{Kalt}},
  \bibinfo{author}{\bibfnamefont{T.}~\bibnamefont{Schimmel}},
  \bibinfo{author}{\bibfnamefont{M.}~\bibnamefont{Wegener}},
  \bibinfo{author}{\bibfnamefont{K.}~\bibnamefont{Okhawa}}, \bibnamefont{and}
  \bibinfo{author}{\bibfnamefont{D.}~\bibnamefont{Hommel}},
  \bibinfo{journal}{Appl.~Phys.~Lett.} \textbf{\bibinfo{volume}{76}},
  \bibinfo{pages}{203} (\bibinfo{year}{2000}).

\bibitem{apl801391}
\bibinfo{author}{\bibfnamefont{H.}~\bibnamefont{Zhao}},
  \bibinfo{author}{\bibfnamefont{S.}~\bibnamefont{Moehl}},
  \bibinfo{author}{\bibfnamefont{S.}~\bibnamefont{Wachter}}, \bibnamefont{and}
  \bibinfo{author}{\bibfnamefont{H.}~\bibnamefont{Kalt}},
  \bibinfo{journal}{Appl.~Phys.~Lett.} \textbf{\bibinfo{volume}{80}},
  \bibinfo{pages}{1391} (\bibinfo{year}{2002}).

\bibitem{s27387}
\bibinfo{author}{\bibfnamefont{D.}~\bibnamefont{Gammon}},
  \bibinfo{author}{\bibfnamefont{E.~S.} \bibnamefont{Snow}},
  \bibinfo{author}{\bibfnamefont{B.~V.} \bibnamefont{Shanabrook}},
  \bibinfo{author}{\bibfnamefont{D.~S.} \bibnamefont{Katzer}},
  \bibnamefont{and} \bibinfo{author}{\bibfnamefont{D.}~\bibnamefont{Park}},
  \bibinfo{journal}{Science} \textbf{\bibinfo{volume}{273}},
  \bibinfo{pages}{87} (\bibinfo{year}{1996}).

\bibitem{jcg742}
\bibinfo{author}{\bibfnamefont{L.}~\bibnamefont{Besombes}},
  \bibinfo{author}{\bibfnamefont{L.}~\bibnamefont{Marsal}},
  \bibinfo{author}{\bibfnamefont{K.}~\bibnamefont{Kheng}},
  \bibinfo{author}{\bibfnamefont{T.}~\bibnamefont{Charvolin}},
  \bibinfo{author}{\bibfnamefont{L.~S.} \bibnamefont{Dang}},
  \bibinfo{author}{\bibfnamefont{A.}~\bibnamefont{Wasiela}}, \bibnamefont{and}
  \bibinfo{author}{\bibfnamefont{H.}~\bibnamefont{Mariette}},
  \bibinfo{journal}{J. Cryst. Growth} \textbf{\bibinfo{volume}{214/215}},
  \bibinfo{pages}{742} (\bibinfo{year}{2000}).

\bibitem{apl3929}
\bibinfo{author}{\bibfnamefont{G.}~\bibnamefont{Eggers}},
  \bibinfo{author}{\bibfnamefont{A.}~\bibnamefont{Rosenberger}},
  \bibinfo{author}{\bibfnamefont{N.}~\bibnamefont{Held}},
  \bibinfo{author}{\bibfnamefont{G.}~\bibnamefont{G{\"u}ntherodt}},
  \bibnamefont{and}
  \bibinfo{author}{\bibfnamefont{P.}~\bibnamefont{Fumagalli}},
  \bibinfo{journal}{Appl. Phys. Lett.} \textbf{\bibinfo{volume}{79}},
  \bibinfo{pages}{3929} (\bibinfo{year}{2001}).

\end{thebibliography}
\end{document}